\documentstyle[aps,prl,multicol,epsf]{revtex}
\tighten

\begin{document}

\title{Temporally disordered granular flow: A model of landslides}

\author{Bosiljka Tadi\'c$^\star $}

\address{
Jo\v{z}ef Stefan Institute,
P.O. Box 3000, 1001-Ljubljana, Slovenia }


\maketitle
\begin{abstract}
\newline
We propose and study numerically a stochastic cellular automaton model
for the dynamics of granular materials with temporal disorder representing
random variation of the diffusion probability $1-\mu (t)$ around
threshold value $1-\mu_0$ during the course of an avalanche.
Combined with the slope threshold dynamics, the temporal disorder yields
a series of secondary instabilities, resembling those in realistic
granular slides. When the parameter $\mu_0$ is lower than the
critical value $\mu_{0}^\star \approx 0.4$, the dynamics is dominated by
occasional huge sandslides.
For the range of values $\mu_{0}^\star \le \mu_0 < 1$ the critical steady
states occur, which are characterized  by multifractal scaling
properties of the slide distributions and
continuously varying critical exponents $\tau _X(\mu_0)$.
The mass distribution exponent for $\mu_0\approx 0.45$ is in
agreement with the reported value that characterizes  Himalayan
sandslides.
At $\mu_{0}=\mu_{0}^\star$ the exponents governing distributions of
large relaxation events reach numerical values which are close to
those of parity-conserving universality class, whereas for small
avalanches they  are close to the mean-field exponents.
\end{abstract}
\pacs{PACS numbers: 64.60.Lx, 05.40.+j,  02.60.Cb, 92.40.Gc }

\begin{multicols}{2}

{\section{Introduction}}

Understanding flow in realistic granular materials appears to be an important
problem from both practical and theoretical point of view \cite{Bak-book},
\cite{PhysTod}.
Renewed theoretical interest in this field has concentrated on the origin
of scaling that characterizes  phenomena in slowly driven granular materials:
Distributions of avalanches in realistic granular piles
\cite{Jaegeretal,Heldetal,RVK,Bretzetal,rice,rice-transport},
stratification\cite{Gene-stratif}, compactification \cite{Coniglio}, etc.
The central question is: Do granular piles self-organize into critical
steady states \cite{Bak-book} and if so, under what conditions?
Another interesting phenomenon related to dynamics of granular materials in
nature is the  landscape evolution due to overland and channel flow, which
results in fractal topography. The underlying mechanisms of erosion with
spatially and temporally varying erosion rates are the subject of intensive
discussion in the literature \cite{erosion}.

It  has been understood that realistic flow in slowly driven granular piles
depends on many parameters, such
as shapes and sizes  (and masses) of individual beans, roughness of
contact surfaces, their wetting properties, etc.
Random (or controlled) variations in some of these parameters lead to
fluctuations of contact angles and force distribution \cite{force-distr},
nonlinear friction,  stochastic character of diffusion, velocity and
convection directions, and  fluctuations in angle of repose.
Unidirectional flow---reflecting dependence on gravity---is common in all
granular materials, as well as occurrence of secondary avalanches following
the initial instability.
Molecular dynamic (MD) simulations \cite{MD} and various cellular automata
models with stochastic relaxation rules \cite{CAM-gran,SCSM1D,LTU} have
been useful in describing certain aspects of realistic granular flow.
However, comparison with measured avalanche properties
 has been only qualitative.

In experiments the most often measured quantity is the outflow current $J$,
which is defined as the number of grains
that leave the system when an avalanche hits its lower boundary.
The probability distribution of outflow current $P(J)$ in the steady
state obeys the scaling form
$P(J,L)=L^{-\beta}{\cal{G}}(JL^{-\nu})$ with $\beta =2\nu $
 when the linear size $L$ of the pile support  is varied, as found
in Ref.\ \cite{Heldetal} for  sandpiles of relatively small sizes.
Using silicon dioxide sand Rosendahl et al. \cite{RVK} concluded that small
and large avalanches behave differently and the distribution $P(J)$ shows no
simple finite-size scaling. Moreover, avalanche statistics was found to
vary with the size of grains used. Measuring the {\it internal} avalanches
Bretz et al. \cite{Bretzetal} have  also observed that two types od
statistics are governing small and large avalanches.
The measured distribution of avalanche size exhibits a power-law behavior
$D(s)\sim s^{-\tau_s}$  with \cite{Bretzetal}
$\tau _s \approx 2.14$, which probably applies for avalanches of small sizes.
The two time (and size) scales were more clearly demonstrated recently by
MD simulations \cite{MD}, leading to two exponents $\tau _s =2$ for short,
and $\tau _s =1.5$ for long time scale.
A sophisticated measurement of the internal avalanches
 was done with a one-dimensional ricepile
\cite{rice}, in which elongated rice grains were used to suppress inertial
effects. Scaling properties of the distribution of dissipated energy
were determined, indicating that details of the dissipation are
responsible for the occurrence of the critical state.
In another experiment the transport of {\it individual} grains was monitored,
and the distribution of transit time was also found to exhibit robust
scaling behavior \cite{rice-transport}.

The collected data for the landslides in nature, triggered by various
mechanisms, also exhibit a power-law behavior \cite{Bak-book}. The
exponents for the area of slides have been estimated in the range
$\tau _s=1.16 - 2.25$ \cite{Turcotte}, depending on the dominating
triggering mechanism and region where the data were collected.
The distribution of the mass collected from Himalayan sandslides
is characterized by the exponent $\tau_m =0.19-0.23$ \cite{Noever}.

In the present work we introduce a new stochastic model of directional
flow on the two-dimensional square lattice in which numerous after-avalanches
are generated within a certain correlation time due to temporal disorder in
the diffusion term. The dynamic
rules are a combination of stochastic diffusion and deterministic branching
processes. The diffusion probabilities change {\it randomly in time},
but are space-independent. Fluctuations in diffusion probability $1-\mu (t)$
around  threshold  value $1-\mu_0$, which depends on external conditions and
thus appears as a control parameter, is motivated by fluctuations in
wetting and drying conditions {\it after an avalanche commenced} (see Sec. II).
Notice that the lifetime of an avalanche can
range from seconds in the laboratory granular piles to geological times
in the landscape evolution. Therefore, the change of local stability
conditions during the avalanche lifetime is a natural choice in the case
of long relaxation times.
A similar type of disorder in directed percolation processes was recently
considered by Jensen \cite{Jensen}.

We perform extensive numerical simulations for various values of the
parameter $\mu_0$ and lattice sizes $L$, and quantify the behavior by
the landslide distributions of: (i) duration $t$---time that an instability
lasts measured on the internal time scale; (ii) size
$s$---area affected by an instability, and  (iii) mass $n$---number
of grains that exhibit slides during one avalanche, and (iv) by outflow
current $J$---number of grains that fall off the open boundaries of the pile.
Self-organized critical states are found for a range of values of the
control parameter $\mu_0\ge \mu_{0}^\star \approx 0.4$, which are
characterized with multifractal scaling properties and $\mu_0$-dependent
critical exponents. For $\mu_0< \mu_{0}^\star $ large
discharging events occur occasionally, representing large-scale erosional
reorganization of the system rather than fluctuations around a well
defined critical state.

The organization of the paper is as follows: In Sec.\ II we introduce
the model and show two representative examples of landslides.
The probability distributions of slides and their scaling properties
are determined in Sec.\ III and IV for various values  of the
linear system size $L$ and the parameter $\mu _0$ in the scaling region.
Sec.\ V contains a short summary and the discussion of the results.

\bigskip

{\section{Model and landslides}}

We consider a square lattice oriented downward, with a dynamic variable,
height $h(i,j)$, associated to each site. The relaxation rules are a
combination of (i) stochastic diffusion by two particles when
$h(i,j)\ge h_c$ with probability $\mu (t)$, which varies in time
(see below); and (ii) deterministic convection, when local slope
$\sigma (i,j)\equiv h(i,j)-h(i+1,j_\pm) $ exceeds some critical value
$\sigma (i,j)\ge \sigma _c$. At each site the rule (ii) is applied
by toppling one particle along an unstable slope repeatedly until both
local slopes drop below $\sigma _c$. The system is updated in parallel,
which leads to a well defined internal time scale of the relaxation
process. The updating is stopped when  {\it all} affected sites become
temporarily stable. Here $(i+1,j_\pm)$ are positions of two downward
neighbors of the site$(i,j)$.
Mass flow is always downward, however, the instability can propagate
backwards both due to nonlocal slope condition and due to time-dependent
diffusion probability. We assume that diffusion probability fluctuates
stochastically in time, but  is space independent. Implementation of
this rule is done as follows: We preset the threshold value $\mu _0$
which is the same for all sites in the system. Then at each site which
is affected by an avalanche a new value $\mu (t)$ is selected at each
time step until the avalanche stops from set of random numbers evenly
distributed on the interval (0,1), and toppling is accepted if
$\mu (t) \le \mu _0$, and rejected otherwise \cite{comment}.
Therefore, for $\mu_0=1$ all sites topple (the rule becomes
deterministic), whereas for $\mu_0 <1$ an unstable site might not
topple at a given time $t$ because of instantly low
diffusion probability $p(t)\equiv 1-\mu (t)<1-\mu_0$, however, it
may topple at a later time step $t^\prime >t$ if  $p(t^\prime )$
exceeds the threshold diffusion probability   $1-\mu_0$.
This temporally varying disorder mimics changes in sticking
properties with time, which then locally influence the angle of repose.
This  phenomenon can be of interest for the flow of granular
materials with large effective friction, such as ricepiles \cite{rice}
in which the effects of granular boundaries may depend on the local dynamic
variable $h(i,j)$ and its derivatives. Therefore the difference
$\mu (t)-\mu_0$ is a measure of the dynamic friction.
Recently proposed models with stochastic critical slope rules in one
dimension  \cite{SCSM1D} proved very successful in describing the
observed {\it transport} properties of ricepiles \cite{rice-transport}.
Whereas for {\it avalanche} distributions these models predict
universal scaling exponents, in contrast to the experimental observations
\cite{Jaegeretal,RVK,Bretzetal,rice}.

Another interesting example is represented by landscape evolution,
which can also be considered as a granular flow\cite{Bak-book},
in which local wetting properties fluctuate in time. By wetting,
$p(t)$ drops below the threshold diffusion probability $1-\mu _0$,
the grains stick together, and the system builds up large local slopes.
At a later time $t^\prime $ these slopes may become
unstable either when due to drying $p(t^\prime )$ exceeds the threshold,
or when the slopes become larger than critical.
Two different classes of triggering mechanisms of landslides have been
discussed in the literature \cite{landslides}: rainfall and water level
which control soil moisture on one side, and ground motion, which leads
to slope variation on the other. The values of measured exponents of
landslide distributions are directly related to the locally prevailing
triggering mechanism \cite{Turcotte}. In principle, threshold shear
stress may depend on the slope angle and on soil properties, which are
influenced by soil moisture. We assume that these two mechanisms
are related  {\it dynamically}. In the present model both mechanisms
are effective: The soil moisture, which affects local height, varies
stochastically in time at each site, whereas we assume that the shear
stress threshold depends only on the local angle and thus remains
deterministic. Moreover, by tuning the critical height mechanism via the
parameter $\mu_0$, we find nonuniversal critical properties and a
transition to noncritical dynamic states, in  qualitative agreement
with experimental observations.
A different model of landslides is obtained by "averaging out" the
critical height mechanism and assuming stochastic variations of critical
slope, which can be viewed as one of few possible  generalizations
of  stochastic critical slope models \cite{SCSM1D} to two dimensions.
So far the results of two-dimensional stochastic critical slope models
are not available in the literature \cite{Al-communication}.

The system is perturbed by adding grains one at a time at a random
site  on the first row, thus increasing local height and slopes. Therefore,
an instability (avalanche) can in principle start only from the top, however,
secondary avalanches are commencing from any affected site in the system,
triggered either by high instant value of $\mu (t)$ or by supercritical slope.
In order to have "clean" statistics, we start each avalanche
from the top row and consider only those secondary avalanches which are {\it
spatially connected } within certain  correlation time $t_c$.
 Here $t_c$ is not a prefixed parameter, but it is determined by
the relaxation process itself. Typically $t_c$ is determined by the lifetime
of the instability, thus  $t_c\gg 1$ for large relaxation events.
There are two interesting limits of our model.  In the limit $\mu_0=1$
it reduces to  the deterministic directed model \cite{DR},
whereas for $\mu_0<1$ and in the limit when the correlation time is
{\it strictly} equal to one, it reduces to the  model considered  in
Ref.\ \cite{LTU}.

The temporally varying diffusion probability is a new ingredient of our
model, which was not considered so far in CA models of granular flow.
It appears to be responsible both for new scaling properties and for
the transition into the state dominated by  large erosional avalanches.
In Fig.\ 1 are shown  two examples of simulated landslides with multiple
topplings due to secondary  avalanches up to fourth degree in the
scaling region (top) and a large erosional event (bottom).

\bigskip
{\section{Probability distributions of slides and their scaling properties}}

In this section we present results of numerical simulations of avalanche
statistics. As discussed in  Sec.\ I, a landslide consists of many
interpenetrating avalanches of different degree, which are
spatially connected one to another within the life-time of the instability.
For concreteness, the probability distributions are determined for the
{\it whole} relaxation event, which is equally termed as avalanche and/or
landslide.
We apply open boundary conditions in the perpendicular direction
(see also later an example where periodic boundaries have been used).
In  most simulations we used $h_c=2$ and $\sigma _c=8$.
By varying the external parameter $\mu_0$ between 0 and 1
and lattice size $L$ between 12 and 192, we determine the distributions
of size, mass and duration of avalanches (slides).

In Figs.\ 2 and 3 the distributions of avalanche duration longer than
$t$,  $P(t)$, size larger than $s$,  $D(s)$,   and mass larger than $n$,
$D(n)$, are shown for $L=128$ and various values of the parameter $\mu_0$.
(Notice that in the deterministic limit $\mu_0=1$
the distributions $D(s)$ and $D(n)$
become identical, however, unbounded number of topplings at each site
for $\mu_0 < 1$ leads to two distinct distributions.)
 For $\mu_0< 1$  a characteristic behavior with two scales appears:
steep section corresponding to small avalanches, and  flat section to
large avalanches. The crossover length between small and large
relaxation events varies with $\mu_0$, however, it remains
small (cf. Figs.\ 2 and 3), so that distributions of avalanches smaller than
the crossover length extend only over one decade.
Here we concentrate on the behavior of large avalanches (i.e.,
avalanches which are larger than the crossover length).
With lowering the threshold diffusion probability $\mu_0$ a large number
of secondary instabilities develop, leading
to the flattening of the distributions. However, we find a power-law
behavior $P(t)\sim t^{1-\tau_t}$, $D(s)\sim s^{1-\tau_s}$, and
$D(n)\sim n^{1-\tau_n}$, as long as $\mu_0\ge 0.4$.
The exponents $\tau_t$, $\tau_s$ and $\tau_n$ appear to vary
continuously with control parameter $\mu_0$, as shown in the insets to
Figs.\ 2 and 3.
The character of the dynamics changes below $\mu_0^\star \approx 0.4$,
where only occasionally very large avalanches occur.
We  study in some more detail the relaxation clusters  at $\mu_0=0.4$.
Numerical values of the exponents are $\tau_t=1.253$,
$\tau _s=1.202$, and $\tau _n=1.190$ for the
distributions of duration, size,  and mass of avalanches, respectively.
In addition, we have measured the distribution of linear elongation of
avalanches in the direction of transport $P(\ell )\sim \ell ^{-\tau _\ell}$,
the mass-to-scale ratio with respect to parallel length $\langle s\rangle _\ell
\sim \ell ^{D_\|}$, and the average transverse extent
$\langle \ell_\bot\rangle \sim \ell ^\zeta $. We find $\tau _\ell =1.578$,
$D_\| =1.572$, and $\zeta =D_\| -1=0.572$ (estimated error bars $\pm 0.03$).
These  values are close to the numerical values of the
exponents  in the parity-conserving universality class \cite{PC}
of branching processes. On the other hand, the exponents governing small
events increase with decreasing $\mu_0$ (cf. Fig.\ 2), reaching the
values $\tau_t^s=$1.92, $\tau_s^s=$ 1.67, and $\tau_n^s=$1.45 for
the duration, size, and mass of small avalanches, respectively, at
$\mu_0=\mu_0^\star$.
Notice that  although the scale of the
the distributions is small, being  bounded by the crossover length,
these values of the exponents indicate closeness of the mean-field
universality class.

\subsection{Multifractal scaling properties of landslide distributions}

By varying the lattice size $L$ with $\mu_0$ fixed in the scaling  region
we study the finite-size effects on the distributions of avalanches.
In contrast to the most of two-dimensional sandpile automata models
in the literature, the present distributions
do not obey simple finite-size scaling. Instead, we
 find that {\it different regions of a large   avalanche have
 different fractal properties} and consequently their  own exponents.
The following multifractal scaling form \cite{mfscal}
\begin{equation}
P(X,L) \sim (L/L_0)^{\phi_X(\alpha _X)} \ ,
\label{MFS}
\end{equation}
with
\begin{equation}
\alpha _X \equiv \left(\log
{{X}\over{X_0}}\right)/\left(\log {{L}\over{L_0}}\right) \
\label{alpha}
\end{equation}
fits well our data with $L_0=1/4$ and $X_0=1/4$. (Here $X\equiv t,s,J$).
In Figs.\ 4 and 5 we show the probability distributions of duration and
size, respectively, for five different lattice sizes $L$ and for
fixed $\mu_0=0.7$ .
The corresponding  spectral functions $\phi _t(\alpha _t )$ vs.
$\alpha _t$  and $\phi _s(\alpha _s )$ vs. $\alpha _s$
are shown in the  insets to  Figs.\ 4 and 5.

{\section{Outflow current}}

The outflow current results only from those avalanches that reach an
open boundary of the system. The size of such events and their
frequency is a relative
measure of the transport processes which occur in the interior of the pile.
The outflow current is easy to measure both in laboratory experiments and
in natural landslides. For instance, the width of the sedimented layers of
granular materials that occur below  steep sections in mountains
are directly related to the size of outflow avalanches from that section.
Sensitivity of the outflow current distribution $P(J)$ to variations in the
control parameter is monitored in our model for $L=48$ with periodic
boundary conditions in the perpendicular direction. In Fig.\ 6 we show
the distribution $P(J)$ vs. $J$ for $\mu_0=$1, 0.8, 0.6., 0.4 and
0.2 . Once again, the change in the character of the
dynamics below $\mu_0^\star$ is also seen in the outflow current, which
becomes centered around a certain mean value (depending on the lattice size).
Above $\mu_0^\star$, we find that the outflow current distribution  exhibits
multifractal scaling properties according to Eqs.\ (\ref{MFS}) and
(\ref{alpha}). The results for $\mu_0=$0.7 and varying $L$ from 12 to 192,
obtained for open boundary conditions in perpendicular direction,
are shown in Fig.\ 7.

Additional information about transport processes in the interior of the
system is obtained by measuring  the outflow current as a function of
time, and time intervals between successive outflow events.
In the inset to Fig.\ 6 we  show the average time
interval between outflow events as a function of the control parameter
$\mu_0$. The time intervals grow exponentially on lowering
$\mu_0$. In Fig.\ 8 the outflow current is shown
as a function of time (measured on the external time scale, i.e., by the
number of added particles), averaged over 1000 time steps for $L=$54 and
with periodic perpendicular boundary conditions. For $\mu_0 \ge 0.4$
(cf. lower three panels), the outflow current fluctuates around mean value
$J_0=1$, thus balancing the input current and maintaining the steady
states of the system (a steady state is characterized by balance between
input and output currents). The amplitude of the outflow events increases
with decreasing $\mu_0$, and at the same time the frequency of events
decreases. This behavior is consistent with the histogram which is shown
in Fig.\ 6. The character of the dynamics changes for $\mu_0 < \mu_0^\star$
(see top panel in Fig.\ 8), with dominating output events of large size
and large time intervals between the events.
At $\mu_0=\mu_0^\star$ a dynamic phase transition occurs between critical
steady states above $\mu_0^\star$ and states without long-range correlations
below $\mu_0=\mu_0^\star$. (Similar phase transitions are found also
in Refs.\ \cite{LTU} and \cite{TD}, however, in different universality
classes.)  Although for $\mu_0 <\mu_0^\star $ the system is likely to
build up a finite slope (unlimited piling is prevented
by the deterministic critical slope rule), preliminary results show that
a substantial growth of the average slope occurs only for  $\mu_0 < 0.2$,
reaching the value $\sigma _c$ at $\mu_0\to 0$.
Further work is necessary in order to investigate the universality class
of this phase transition.

{\section{Discussion and Conclusions}}
In the present model, combined relaxation rules with temporal disorder are
responsible for numerous after-avalanches, which lead to large relaxation
events resembling sandslides  in realistic granular materials.
Numerical simulations show that such large relaxation events
exhibit scaling behavior for a range of values of the the control
parameter $\mu _0\ge \mu _0^\star \approx 0.4$. The avalanche
distributions are characterized by  continuously
varying scaling exponents, in  qualitative agreement with the
data collected from natural landslides. Moreover,  comparison
of the exponent of the avalanche mass distribution  $\tau _n$
for $0.4< \mu _0< 0.5$ with the one that characterizes Himalayan
sandslides reported in Ref.\ \cite{Noever} is satisfactory.
For various lattice  sizes the distributions are characterized by
multifractal rather than finite-size scaling properties.
The deterministic part of the relaxation rules leads  to branching
processes with, on the average, even number of offsprings.
For this reason  the scaling exponents for the distributions reach
numerical values characteristic of the modulo-two conserving
processes (also known as parity conserving processes) before scaling
behavior  disappears at $\mu _0= \mu _0^\star $.
Below $\mu _0^\star $ the {\it critical} steady state is lost. The
dynamics is dominated by large erosional avalanches in a region close
to $\mu_0^\star$ and a net average slope appears for smaller
values of $\mu_0$.

\section*{Acknowledgments}

This work was supported by the Ministry of Science and Technology of the
Republic of Slovenia. I would like to thank R. Pastor-Satorras, A. Corral,
and D.L. Turcotte for helpful discussions.


\narrowtext
\begin{figure}[thb]
\epsfxsize=82mm\epsffile[18 245 592 594]{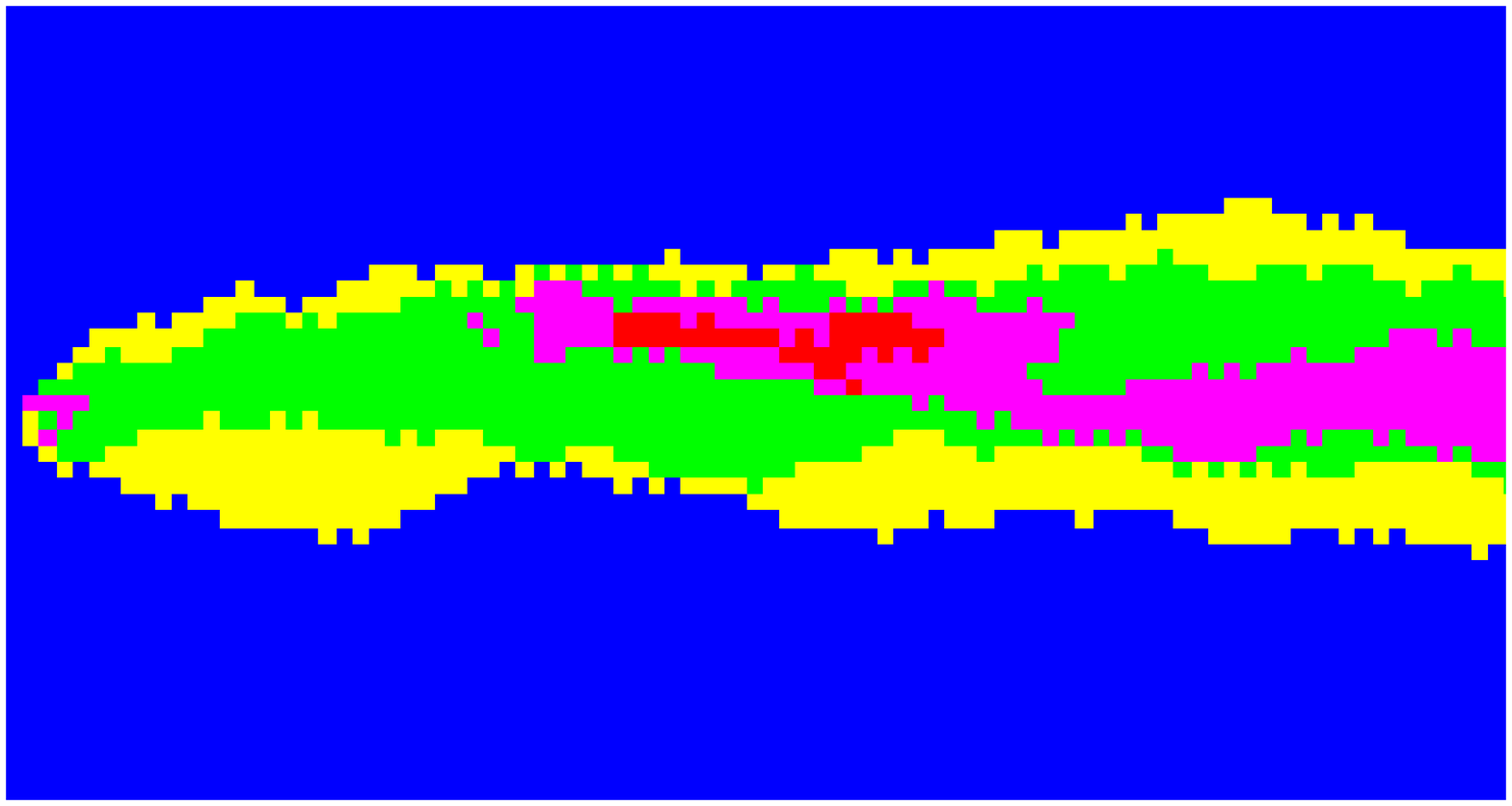}

\epsfxsize=82mm\epsffile[18 234 592 556]{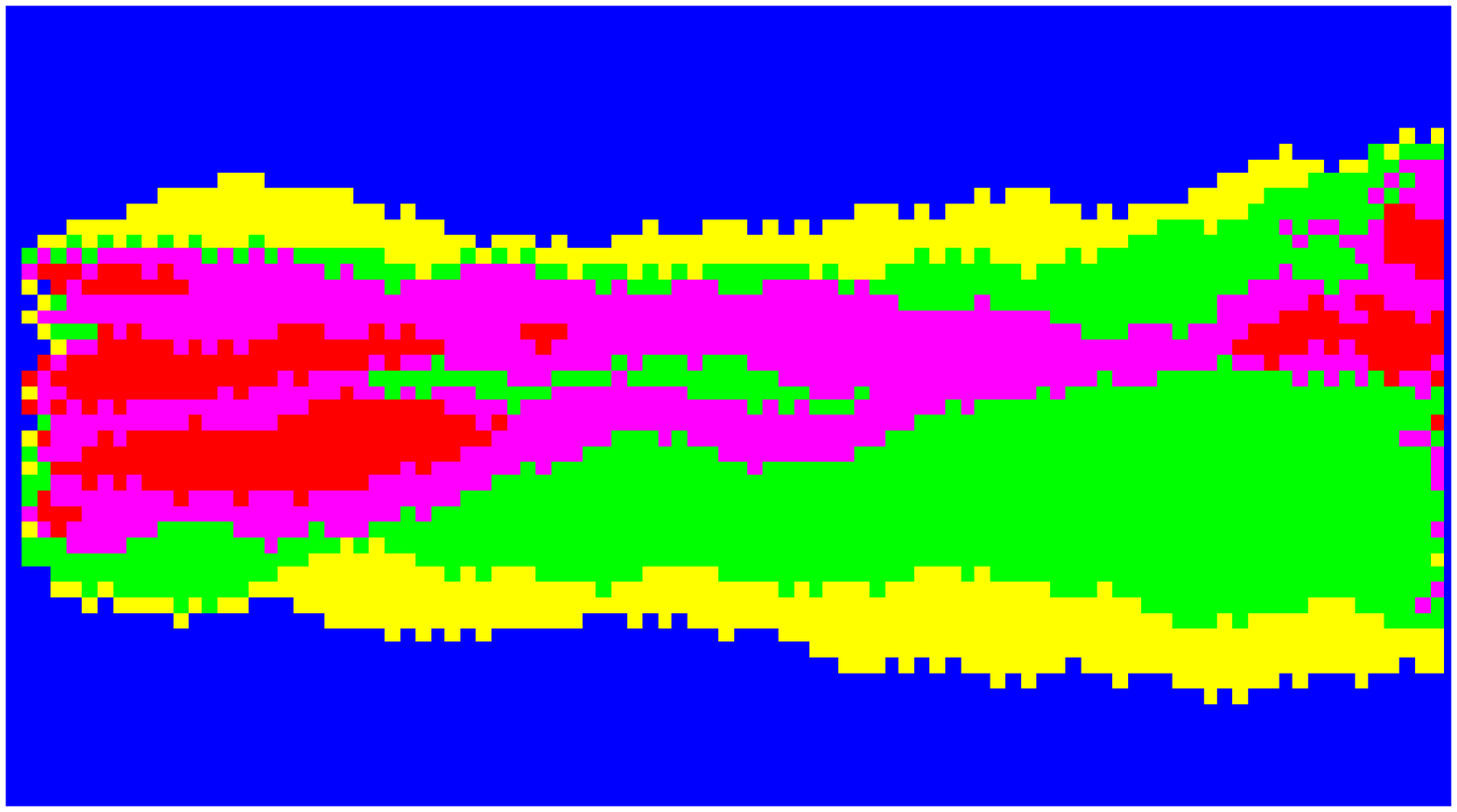}
\caption{\label{fig1} Two examples of  avalanches running from left
to right: (top) in the scaling region $\mu_0>\mu_0^\star$ and
(below) in the region of erosional avalanches. Multiple topplings up
to forth order are marked by different degrees of gray color.}
\end{figure}

\begin{figure}[thb]
\epsfxsize=82mm\epsffile[28 65 563 562]{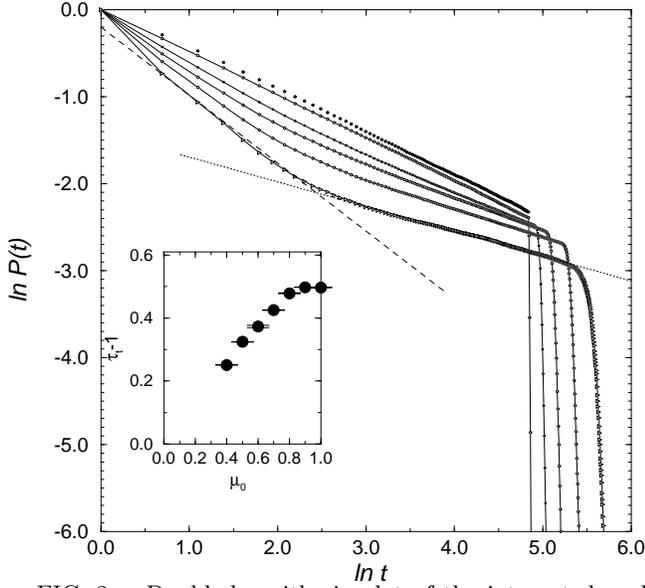}
\caption{\label{fig2} Double-logarithmic plot of the integrated
probability distribution of avalanche durations $P(t)$ vs. duration $t$
for $L=128$ and various values of the control parameter $\mu_0$=1,
0.9, 0.8, 0.7, 0.6, and 0.5 (top to bottom). Dashed and dotted lines
indicate slopes of small and large avalanches, respectively.
Inset: Scaling exponent of
large avalanches $\tau_t-1$ vs. $\mu _0$.}
\end{figure}

\begin{figure}[thb]
\epsfxsize=82mm\epsffile[37 66 508 751]{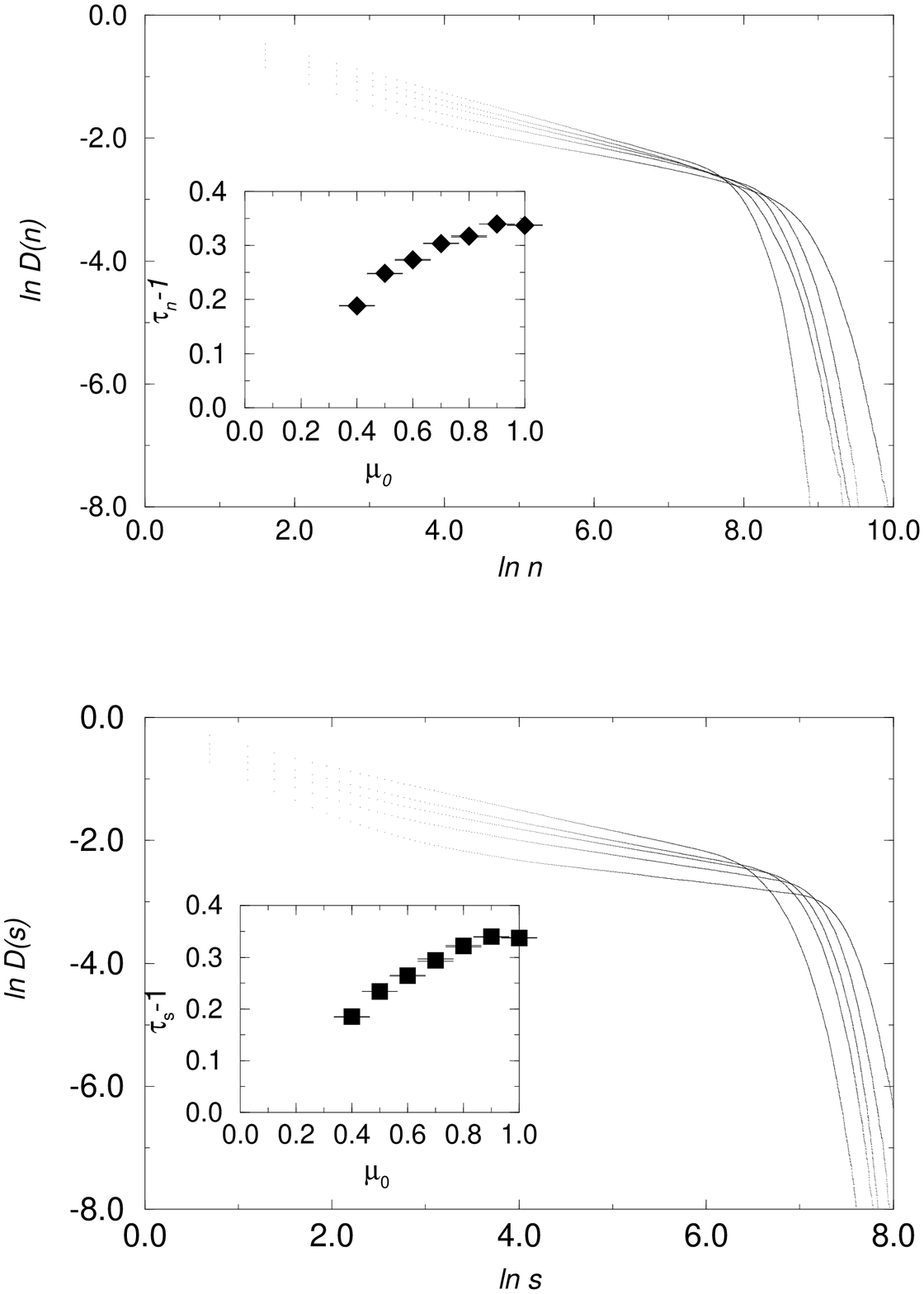}
\caption{\label{fig3} Double-logarithmic plot of the integrated
probability distribution of avalanche size $D(s)$ vs. size $s$
(bottom) and mass $D(n)$ vs. $n$ (top), for $L=128$ and for (top to bottom)
$\mu_0$=1, 0.8, 0.7, 0.6, and 0.5 .
Inset: Scaling exponent of large avalanches $\tau_s-1$ vs. $\mu_0$
(bottom figure) and  $\tau _n-1$ vs. $\mu_0$ (top figure).}
\end{figure}

\begin{figure}[thb]
\epsfxsize=82mm\epsffile[28 65 563 562]{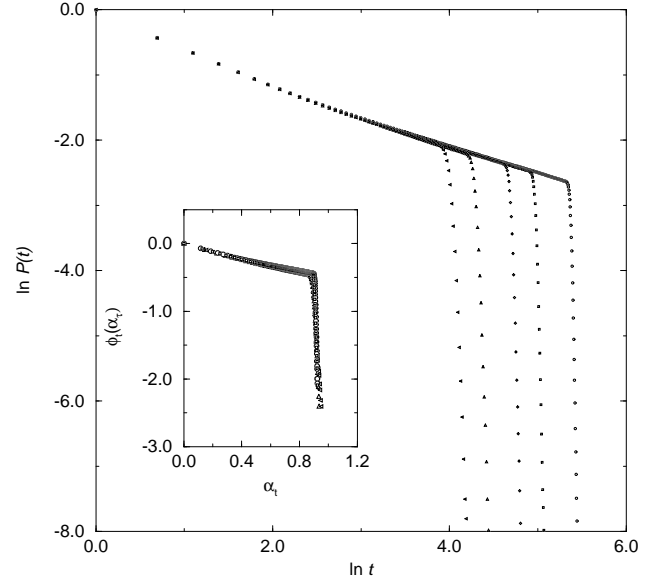}
\caption{\label{fig4} Double-logarithmic plot of the distribution $P(t)$
vs. $t$ for $\mu_0$=0.7 and for various lattice sizes $L$=12, 24, 48, 96, and
192 (left to right) with open boundary conditions.
Inset: Multifractal scaling function $\phi _t(\alpha _t)$
vs. $\alpha _t$ .}
\end{figure}

\begin{figure}[thb]
\epsfxsize=82mm\epsffile[28 65 563 562]{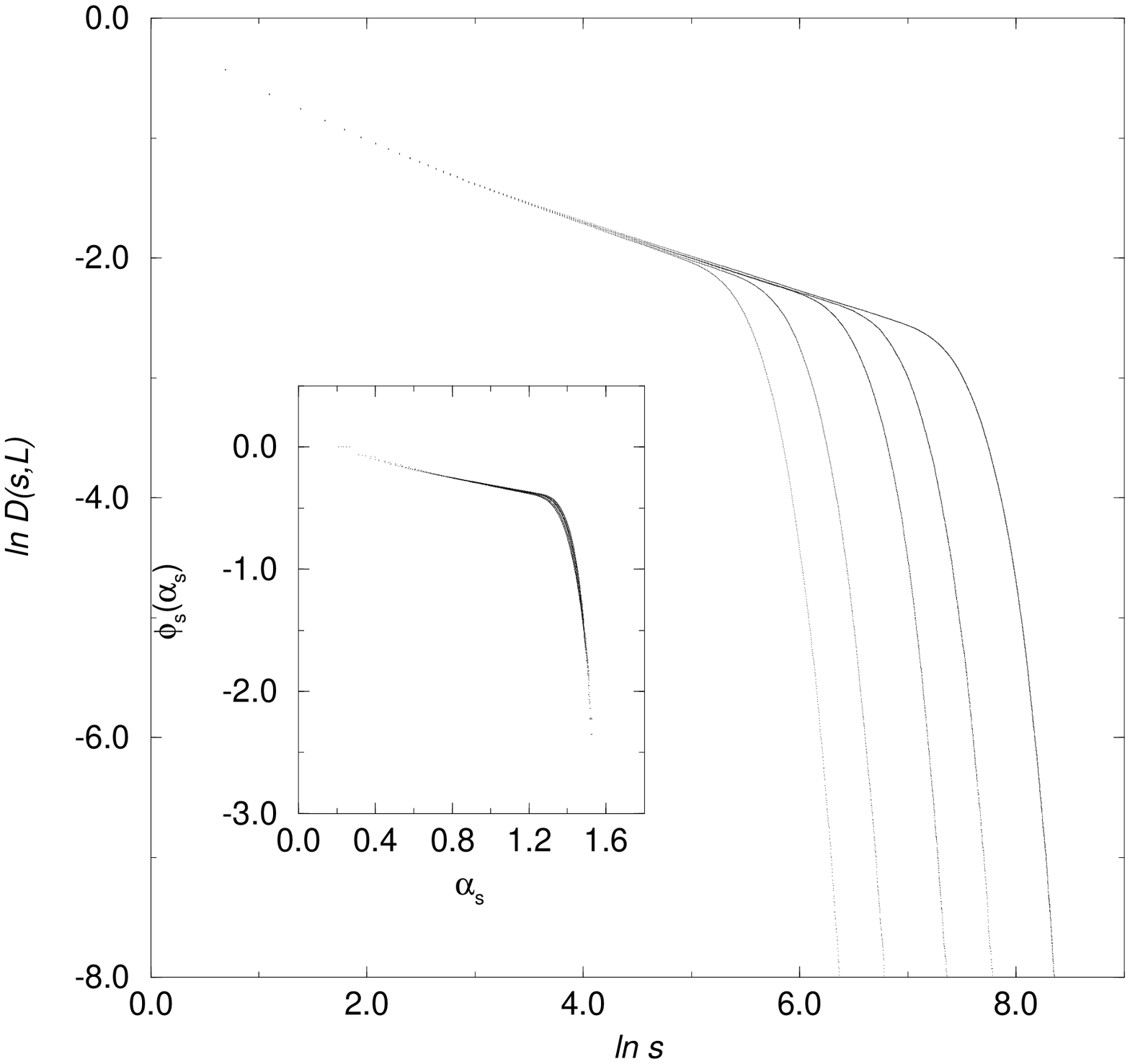}
\caption{\label{fig5}
Double-logarithmic plot of the distribution
of sizes $D(s)$ vs. $s$ for the same set of parameters as Fig.\ 4.
Inset: Multifractal scaling function
$\phi _s(\alpha _s)$ vs. $\alpha _s$ .}
\end{figure}

\begin{figure}[thb]
\epsfxsize=82mm\epsffile[28 65 563 562]{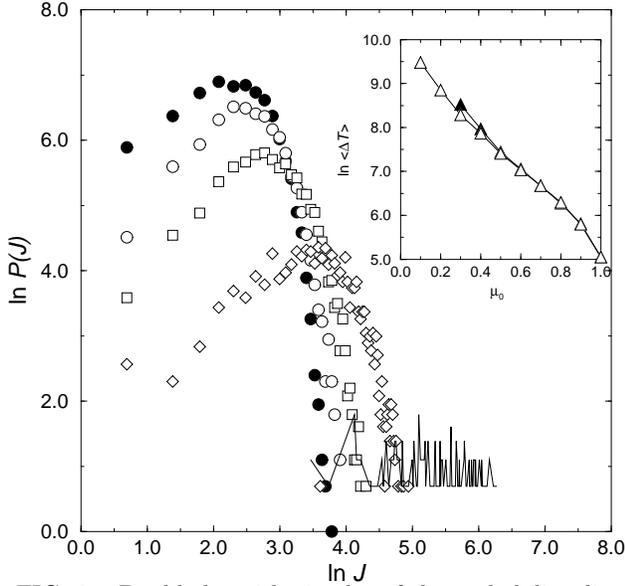}
\caption{\label{fig6} Double-logarithmic plot of the probability
distribution of outflow current $P(J)$ vs. $J$ for $L$=48 with
periodic boundary conditions in perpendicular direction,
and for $\mu_0$=1, 0.8, 0.6, 0.4, and 0.2 (top to bottom).
Inset: Average time intervals between outflow events
on the same lattice for $\sigma _c$=8 (open triangles) and $\sigma _c$=4
(filled triangles) .}
\end{figure}
\begin{figure}[thb]
\epsfxsize=82mm\epsffile[28 65 563 562]{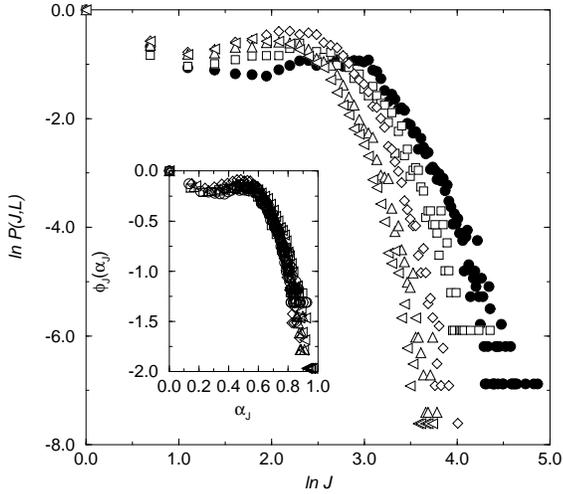}
\caption{\label{fig7} Distribution of outflow current measured with open
boundary conditions  for various lattice sizes $L$=12, 24, 48, 96, and 192
(left to right) and for fixed $\mu_0$=0.7.
Inset: Spectral function $\phi _J(\alpha _J)$ vs. $\alpha _J$ .}
\end{figure}

\begin{figure}[thb]
\epsfxsize=82mm\epsffile[70 68 508 714]{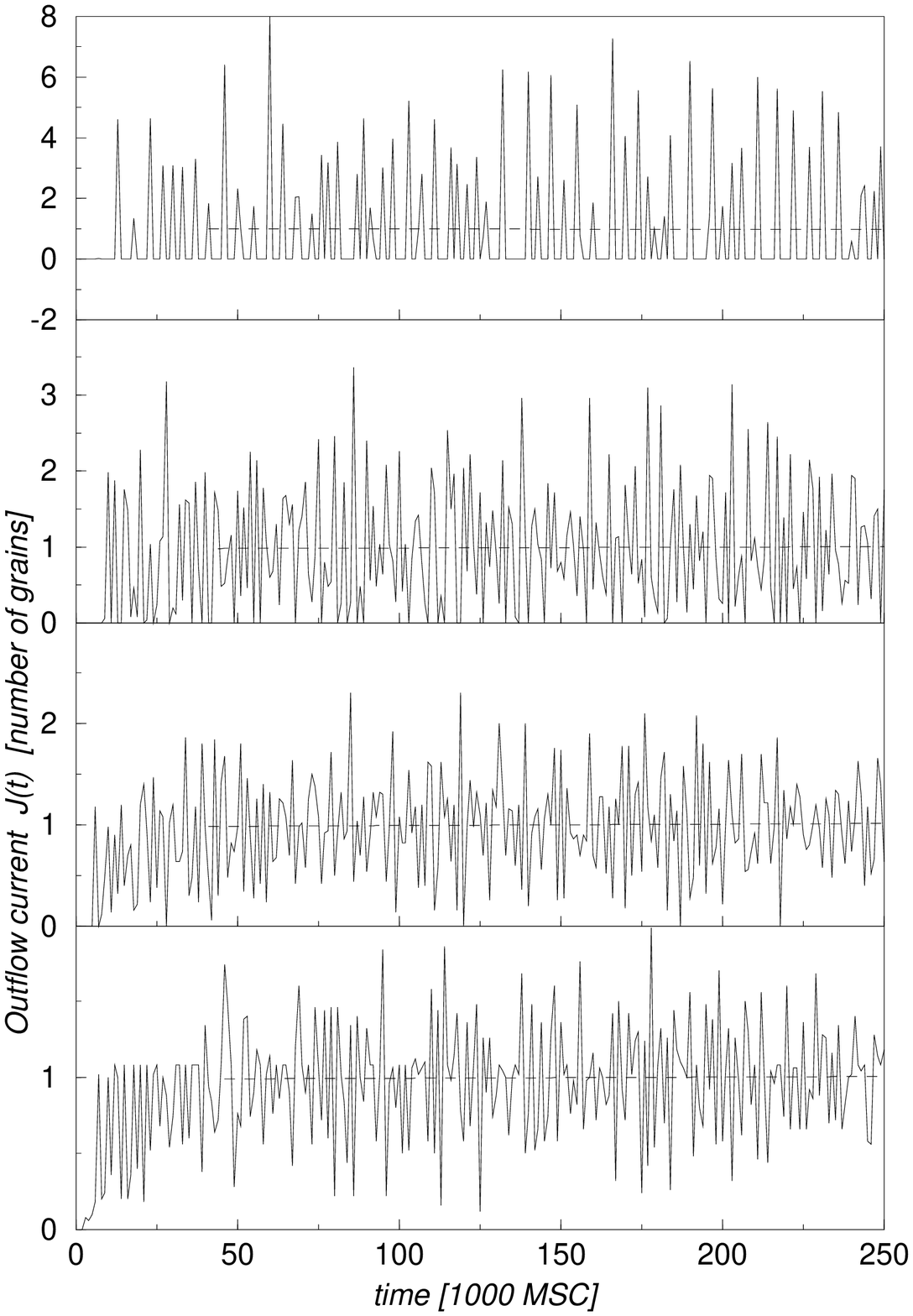}
\caption{\label{fig8} Outflow current $J(t)$ vs. time $t$ (measured
in number of added particles), averaged over 1000 time steps, for
$L$=54 and $\mu_0$=1, 0.7, 0.4, and 0.3 (bottom to top) and
with periodic boundary conditions. Dashed lines are mean values calculated
by linear fits of the data for $t>40$: (bottom to top) 0.9972, 1.0004,
0.9935 and 0.9892.
Slopes of the dashed lines are smaller than $10^{-5}$ in each case.}
\end{figure}

\end{multicols}


\begin{references}
\bibitem[18]{Noever} D.A. Noever, Phys. Rev. E {\bf 47}, 724 (1993).
Notice that the original data were collected by clearing the mountain
roads, indicating that the reported exponent applies for the integrated
probability distribution of mass.
\bibitem[20]{comment} Notice that in this implementation the width of
the distribution of $\mu (t)$ does not appear as a tunable parameter.
\end{references}
\end{document}